\documentstyle[twoside,fleqn,espcrc2,epsfig]{article}

\newcommand{\be}{\begin{equation}}
\newcommand{\ee}{\end{equation}}
\newcommand{\ba}{\begin{eqnarray}}
\newcommand{\ea}{\end{eqnarray}}

\newcommand{\AmS}{{\protect\the\textfont2
  A\kern-.1667em\lower.5ex\hbox{M}\kern-.125emS}}

\title{On $K\to\pi\pi$ Decays in Quenched and Unquenched Chiral 
Perturbation Theory} 

\author{ Maarten Golterman\address{Department of Physics, Washington
University, St. Louis, MO 63130, USA}
and 
Elisabetta Pallante\address{%
Facultat de F\'\i sica, Universitat de Barcelona,
 Diagonal 647, 08028 Barcelona, Spain}%
\thanks{Presented by E. Pallante} }
       
\begin{document}

\begin{abstract}

We calculate the logarithmic corrections to the matrix elements for
$K^+\to\pi^+$ and $K\to\ {\rm vacuum}$
(which are used on the lattice to determine $K\to\pi\pi$ amplitudes), 
in one-loop quenched and
unquenched Chiral Perturbation Theory.  We find that these corrections
can be large.  We also discuss, and present some results for, the
direct determination of $K\to\pi\pi$ amplitudes.  In particular,
we address effects from choosing $m_s=m_d$ and vanishing external
spatial momenta, finite volume and quenching. 
In the quenched octet case, we find {\em enhanced}
finite-volume contributions which may make numerical estimates of
this matrix element unreliable for {\em large} volumes.
\end{abstract}

\maketitle

\section{Introduction}

Chiral Perturbation Theory (ChPT) helps us understand several systematic 
errors which afflict lattice computations of $K\to\pi\pi$ decay amplitudes,
and thus plays an important role in assessing the reliability of such
computations.  In particular, ChPT can be used to gain insight into the
size of finite-volume and quenching effects, as well as 
the modifications induced by an 
unphysical choice of kinematics and$/$or the values of light quark masses.
We consider two approaches to these amplitudes. The first is the direct
computation of such amplitudes with $m_s =m_d$ and external
mesons at rest \cite{BERNARD}. Three key questions can be studied in ChPT: 
1) how much do these unphysical choices affect the size of the chiral 
logarithms (with and without quenching)?; 2) are there quenched chiral 
logarithms \cite{qCHPT}?; 
3) are there {\em enhanced} finite-volume corrections?
Here we mainly answer the last two questions, while a more detailed
analysis will be given elsewhere \cite{EG}. 
Second, we calculate the $K\to \pi$ and
$K\to 0$ ($K$ to vacuum) matrix elements at one loop in ChPT, 
unquenched and quenched.
The motivation comes from the possibility of performing 
an indirect 
determination of $K\to \pi\pi$ amplitudes through the 
computation of reduced matrix elements such as 
$K\to \pi$ and $K\to 0$, which is simpler on the lattice \cite{KP}.

\section{The Unphysical $K^0\to \pi^+\pi^-$ amplitude}

The Euclidean effective Lagrangian for $\Delta S =1$ hadronic weak 
transitions can,
at leading order in ChPT, be written as \cite{KP,EP1} (notation of
\cite{KP}):
\ba
{\cal L}_{{\tiny{\Delta S=1}}}&=& 
- \alpha^{27}T^{ij}_{kl}(\Sigma\partial_\mu\Sigma^\dagger)^k_{\ i}
(\Sigma\partial_\mu\Sigma^\dagger)^l_{\ j}
\label{WEAK}\\
&&\hspace{-2truecm}-\alpha^8_1
\mbox{tr}[\Lambda(\partial_\mu\Sigma)(\partial_\mu\Sigma^\dagger)]
\!+\!\alpha^8_2\frac{8v}{f^2}\mbox{tr}
[\Lambda(\Sigma M\!+\!M^\dagger\Sigma^\dagger)]
\;,
\nonumber
\ea
where the first term transforms as $(27_L,1_R)$ under SU(3)$\times$SU(3)
and the last two terms as $(8_L,1_R)$.

\begin{figure}
\begin{center}
\leavevmode\epsfxsize=7cm\epsfbox{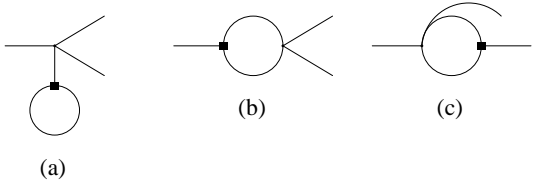}
\end{center}
\vskip -0.7cm
\caption{Some $K\to\pi\pi$ diagrams in ChPT. 
The box is a weak vertex, the dot a strong 
vertex.
\label{TADPOLE}}
\vskip -0.7cm
\end{figure}

The term with coupling $\alpha^8_2$ is known as the ``weak mass term,"
and mediates the $K\to 0$ transition at tree level. Its odd-parity 
part, which in principle can also
contribute to the octet $K\to \pi\pi$ amplitude, 
is proportional to $m_s -m_d$. For $m_s\neq m_d$ the weak mass term is a total 
derivative \cite{KP,BPP}, and therefore does not contribute to any physical 
matrix element. Whether this term contributes to the octet 
$K\to \pi\pi$ matrix element for unphysical external momenta and $m_s =m_d$ 
is a more subtle question.
What actually is computed on the lattice is the Euclidean 
correlation function $C(t_1,t_2)=\langle 0\vert \pi^+(t_2)\pi^-(t_2) O_8(t_1)
\overline{K}^0(0)\vert 0\rangle$.
Any contribution generated by the insertion of the
weak mass term to the Euclidean 
correlation function at fixed times is proportional to $m_s -m_d$ in the limit 
$m_s\to m_d$ and therefore zero at $m_s =m_d$; there are no subtleties
with propagator poles in
Euclidean space from tree-level tadpole diagrams \cite{EG}. 
This is also true for tadpole contributions as in diagram (a) of  
Fig. \ref{TADPOLE} with the insertion of an octet or a 27-plet weak operator.
Such contributions are absent for $m_s =m_d$.
We note that, choosing quark masses such that 
$m_K = 2m_\pi$, as proposed in \cite{MARTI}, the contribution from 
$\alpha^8_2$ vanishes for the same reason as for the 
physical $K\to \pi\pi$ amplitude, but that, in general, 
there are contributions from Fig. \ref{TADPOLE}(a). 

While the unphysical choice of kinematics and quark 
masses modifies the size of chiral logarithms and 
finite-volume corrections, quenching also causes new
``quenched-artifact" contributions, due to the presence of the double 
pole in the singlet propagator.
These are of two types: quenched chiral logarithms (Q$\chi$L) and {\em enhanced}
finite-volume corrections (discovered in quenched 
 $\pi\pi$ scattering \cite{BG}). In principle,
both artifacts occur in the quenched octet $K\to \pi\pi$ amplitude.

Only diagrams  of type (b), (c) in Fig. \ref{TADPOLE} 
with the weak operator $\alpha^8_1$ 
give rise to Q$\chi$L in the unphysical  $K^0\to \pi^+\pi^-$ amplitude
at $m_s =m_d$. However, we find that {\em no}  Q$\chi$L is present at one loop,
due to a cancellation between contributions from type-(b) and type-(c)
diagrams. 

In finite volume, only the ``rescattering diagram" of type (b) 
gives rise to power-like finite-volume corrections.  
It was shown in the case of $\pi\pi$ scattering \cite{BG} how, in the 
quenched approximation in a similar diagram, the presence of a double-pole
singlet propagator gives rise to enhanced (infrared divergent!)
finite-volume corrections. The same happens for the  
octet $K^0\to\pi^+\pi^-$ amplitude. 

We have calculated the chiral logs and power-like finite-volume
corrections for $C(t_1,t_2)_{\mbox{octet}}$ to one loop.  
In the unquenched case we find
\ba
C(t_1,t_2)&=& {8i\alpha^8_1\over f^3} 
{M^2L^3 \over (2 M)^3} e^{-2M (t_2-t_1)-Mt_1} \nonumber\\
&& \hspace{-1.0truecm}   
\left [ 1-\mu(M)
+{7\over 6}{1\over{f^2L^3}}(t_2-t_1) \right.\nonumber\\
&& \hspace{-1.0truecm}   \left.
-{{M^2}\over{(4\pi f)^2}}
\left({41.597 \over{ML}}+
{{62}\over 3}{{\pi^2}\over{(ML)^3}}\right)\right]\, , \label{UR}
\ea
where $M$ is the degenerate meson mass, $f$ is the pion 
decay constant in the chiral limit (normalized such that its value is 132\ MeV
at the physical pion mass),
and $\mu (M) =(M^2/(16\pi^2f^2))\log (M^2/\Lambda^2)$ is the chiral logarithm.
In the quenched case we obtain 
\ba
C(t_1,t_2)&=& {8i\alpha^8_1\over f^3}
{M^2 L^3 \over (2 M)^3} e^{-2M (t_2-t_1)-Mt_1}
 \label{QR}\\
&& \hspace{-2.0truecm}   
\left[1
+\delta\left(-{{\pi^2}\over{M^2L^3}}(t_2-t_1)
-{{2\pi^2}\over{ML^3}}(t_2-t_1)^2
  \right. \right.\nonumber\\
&& \hspace{-2.0truecm}   
\left. 
+{{3\pi^2}\over{4(ML)^3}}+{2.2284\over{ML}}
-0.41877ML\!\right)\!+\! 2\alpha\mu(M)\nonumber\\
&&\hspace{-2.0truecm}
 +{\alpha\over 3}{{M^2}\over{(4\pi f_\pi)^2}}\left(
{{5\pi^2}\over{2(ML)^3}} -{{14\pi^2}\over{M^2L^3}}(t_2-t_1)
\right.\nonumber\\
&&\hspace{-2.0truecm}
\left.\left.
+{{4\pi^2}\over{ML^3}}(t_2-t_1)^2
+{{31.198}\over{ML}}  +0.83754ML\right)
\right] , \nonumber
\ea
where $\delta =m_0^2/(24\pi^2f^2)$ contains the singlet mass $m_0$ 
and $\alpha$ is another singlet parameter renormalizing its kinetic term
\cite{qCHPT}.

We have ignored $O(p^4)$ contact terms, exponentially suppressed
finite-volume effects, and contributions from excited states.  The term
linear in $t_2-t_1$ can be related to finite-volume energy shifts of
the two-particle internal states of type-(b) diagrams (at least in the
unquenched case).  The term linear in $ML$ inside the square brackets
of Eq. (\ref{QR}) is the enhanced finite-volume contribution, which is a
quenched artifact, as is the term quadratic in $t_2-t_1$.

\section{$K^+\to \pi^+$ and $K\to 0$ matrix elements }

Ref. \cite{KP} proposed an indirect determination of  
the  $K\to \pi\pi$ amplitudes with $\Delta I = 1/2$ and $3/2$ by computing on 
the lattice the reduced amplitudes  $K^+\to \pi^+$ and $K\to 0$.
In particular, at tree level in ChPT, the ratio of the  
$\Delta I = 1/2$ and $3/2$
 $K\to \pi\pi$ amplitudes can be determined through 
\be 
{ [K^0\!\to\!\pi^+\pi^-]_{\tiny\frac{1}{2}} \over [K^0\!\to\!\pi^+\pi^-]_{\tiny
\frac{3}{2}} } =
{  [K^+\!\to\!\pi^+]_{\tiny\frac{1}{2}} - b [K^0\!\to\! 0]_{\tiny\frac{1}{2}} 
\over [K^+\!\to\!\pi^+]_{\tiny\frac{3}{2}} }\, ,
\label{RATIO}
\ee
where $b = iM^2/f(m_K^2-m_\pi^2)$, $M$ is the degenerate mass 
used to compute $K^+\to \pi^+$, and $m_K,\, m_\pi$
are the nondegenerate masses used to compute $K\to 0$.
The question arises how one-loop corrections 
modify Eq. (\ref{RATIO}). This problem was already addressed in 
\cite{BPP} in the unquenched case, however, what is calculated 
there is the full 
pseudoscalar two-point function, and not the amplitude $K\to \pi$.

Here, we present the chiral logs for $K^+\to\pi^+$ and $K\to
0$. For $K^+\to\pi^+$, with degenerate masses,
we find for $\Delta I=1/2$, unquenched,
\ba
\frac{[K^+\to\pi^+]}{4M^2/f^2}&=&
\alpha^8_1\left( 1- {1\over 3} \mu (M)\right )
 \nonumber\\
&& \hspace{-2.5truecm}  
-\alpha^8_2  \left( 1 +\frac{4}{3} \mu (M)\right ) 
- \alpha^{27} \left( 1 -12 \mu (M)\right )\, ,
\ea
while in the quenched case we obtain
\ba
\frac{[K^+\to\pi^+]}{4M^2/f^2}&\!\!\!=&
\!\!\!\alpha^8_1 \left( 1 -2\delta\log \frac{M^2}{\Lambda^2} +4\alpha
 \mu (M)\right ) \nonumber\\
&& \hspace{-2.3truecm} 
-\alpha^8_2  \left( 1 +{4\over 3}\alpha\mu (M)\right ) 
- \alpha^{27} \left( 1 -6 \mu (M)\right )\, .
\ea
The $\Delta I=3/2$ amplitudes are obtained from this by setting $\alpha^8_1
=\alpha^8_2=0$.
Note that the contribution from ordinary chiral logarithms is 
substantially reduced by quenching.  One should keep in mind that
the values of the $\alpha$'s are in principle different in the
quenched and unquenched theories.

The one-loop $K\to 0$ amplitude to leading order 
in $m_K^2 -m_\pi^2$ is (with $M^2$ some average of $m_K^2$ and $m_\pi^2)$), 
unquenched, 
\ba
\frac{[K\to 0]f}{4(m_K^2\!-\!m_\pi^2)}&\!\!\!=\!\!\!&\!i\alpha^8_2\! \left (
\!1\!-\!{13\over 3} \mu (M)\!\right )\! 
+\!i\alpha^8_1{10\over 3} \mu (M)\, ,\nonumber
\ea
while the quenched amplitude is
\ba
\frac{[K\to 0]f}{4(m_K^2\!-\!m_\pi^2)}&\!\!\!=\!\!\!&\!i\alpha^8_2\! 
+\!i \alpha^8_1\! \left (\!2\delta  \log \frac{M^2}{\Lambda^2} 
-4\alpha \mu (M)\!\right )\!.\nonumber
\ea
Again, the chiral logarithms are potentially large, and
 reduced by quenching. One can now in principle extract unquenched
\ba
\frac{\alpha^8_1}{\alpha^{27}}\!=\!
\frac{[K^+\to\pi^+]_8(1-3\mu)\!-\!b[K\to 0](1+\frac{8}{3}\mu)}
{-[K^+\to\pi^+]_{27}(1+12\mu)} , \nonumber
\ea 
and quenched
\ba
\frac{\alpha^8_1}{\alpha^{27}}=
\frac{[K^+\to\pi^+]_8-b[K\to 0](1+\frac{4}{3}\alpha\mu)}
{-[K^+\to\pi^+]_{27}(1+6\mu)} \, , \nonumber
\ea 
where $\mu\equiv\mu(M)=(M^2/(16\pi^2f^2))\log (M^2/\Lambda^2)$.
It is clear that one-loop corrections are potentially large in the 
determination of weak-Lagrangian parameters from lattice computations.
(For $M=400$ MeV, $\Lambda=m_\rho$ and $f=132$ MeV, $\mu(M)=-0.076$,
and {\it e.g.} $1+6\mu=0.54$.)

Obviously, in order to go beyond these ``leading-log" estimates, 
 it is necessary to consider also the contributions  of $O(p^4)$ LECs 
to $K^+\to\pi^+$ and $K\to 0$.
Then, it is worth looking for ratios less sensitive to one-loop effects, 
if such exist (considering also other channels
like  $K\to\eta$ and/or varying momenta or masses), and also, 
whether (combinations of) the $O(p^4)$ 
LECs that appear in $K\to\pi\pi$ amplitudes
can be extracted from amplitudes with less external legs.

\bigskip
We thank Claude Bernard and Steve Sharpe for very useful discussions.
MG is supported by the US Dept. of Energy, and EP by the Minist. de Educacion 
y Cultura (Spain).


\end{document}